\documentstyle[11pt,iau185,twoside,epsf]{article}

\newcommand{\ms}{\mbox{m\,s$^{-1}$}}
\newcommand{\bhyi}{\object{$\beta$~Hyi}}
\newcommand{\muHz}{\mbox{$\mu$Hz}}

\begin{document}
\title{Solar-Like Oscillations in $\beta$ Hydri: Evidence for Short-Lived
High-Amplitude Oscillations}
\author{
T.R. Bedding$^1$,
H. Kjeldsen$^2$,
I.K. Baldry$^3$,
F. Bouchy$^4$,
R.P. Butler$^5$,\\
F. Carrier$^4$,
F. Kienzle$^4$, 
G.W. Marcy$^6$,
S.J. O'Toole$^1$,
C.G. Tinney$^3$
}

\affil{
$^1${School of Physics A28, University of Sydney, NSW 2006,
Australia.}
\\
$^2${Theoretical Astrophysics Center, University of Aarhus, Denmark.}
\\
$^3${Anglo-Australian Observatory, Epping, Australia.}
\\
$^4${Observatoire de Gen\`eve, Sauverny, Switzerland.}
\\
$^5${Carnegie Institution of Washington, Washington, DC, USA. }
\\
$^6${Department of Astronomy, University of California, Berkeley, CA.}
}

\begin{abstract}
Velocity measurements of the G2 subgiant \bhyi{} with both UCLES and
CORALIE show a clear excess of power centred at 1.0\,mHz.  In the UCLES
data we find evidence for a short-lived, high-amplitude oscillation event.
If confirmed as a feature of subgiants, such `starquakes' would make it
harder to measure accurate mode frequencies and perform asteroseismology.
\end{abstract}

 
\section{Introduction}

As reported by Bedding et al.\ (2001) and Carrier et al.\ (2001), we have
made a clear detection of excess power from Doppler observations of the G2
subgiant \bhyi.  We observed this star with the UCLES echelle spectrograph
on the 3.9-m Anglo-Australian Telescope, using an iodine absorption cell as
a velocity reference, and also with the CORALIE spectrograph on the 1.2-m
Leonard Euler Swiss telescope at La Silla.  In both sets of data we see a
clear excess of power centred at 1.0\,mHz, with peak amplitudes of about
0.5\,\ms, in agreement with expectations for this star.  The time series of
velocity measurements from UCLES is shown in Fig.\,1.
\begin{figure}
\begin{center}
\mbox{\epsfxsize=0.95\textwidth\epsfbox{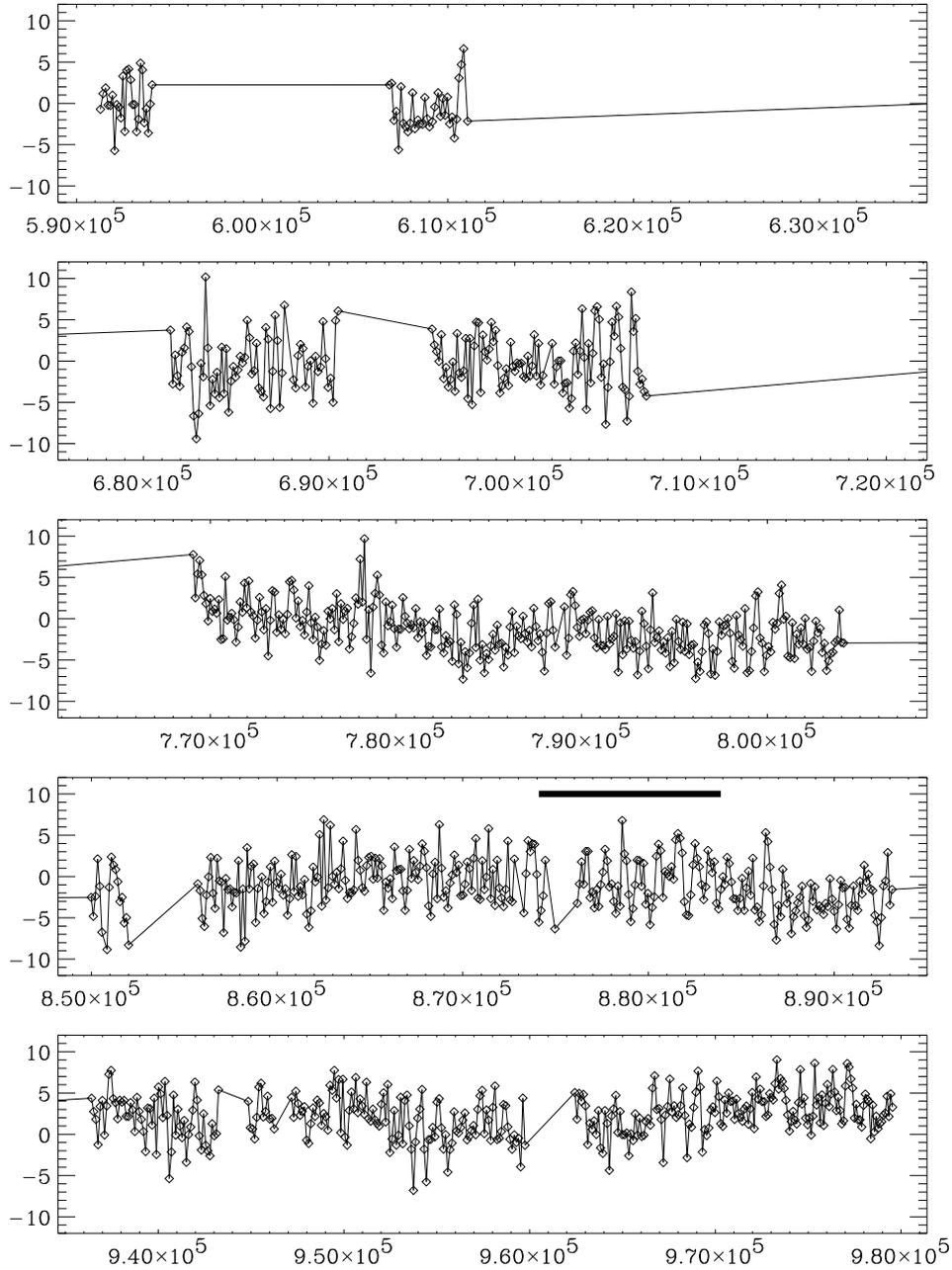}}
\caption{Velocity measurements in m\,s$^{-1}$ from the UCLES
observations.  Time is measured in seconds since JD~2451700 and each panel spans
13 hours.  The horizontal line on the fourth night is discussed in Sect.\,3. }
\end{center}
\end{figure}
\begin{figure}[h!]
\begin{center}
\mbox{\epsfxsize=0.95\textwidth\epsfbox{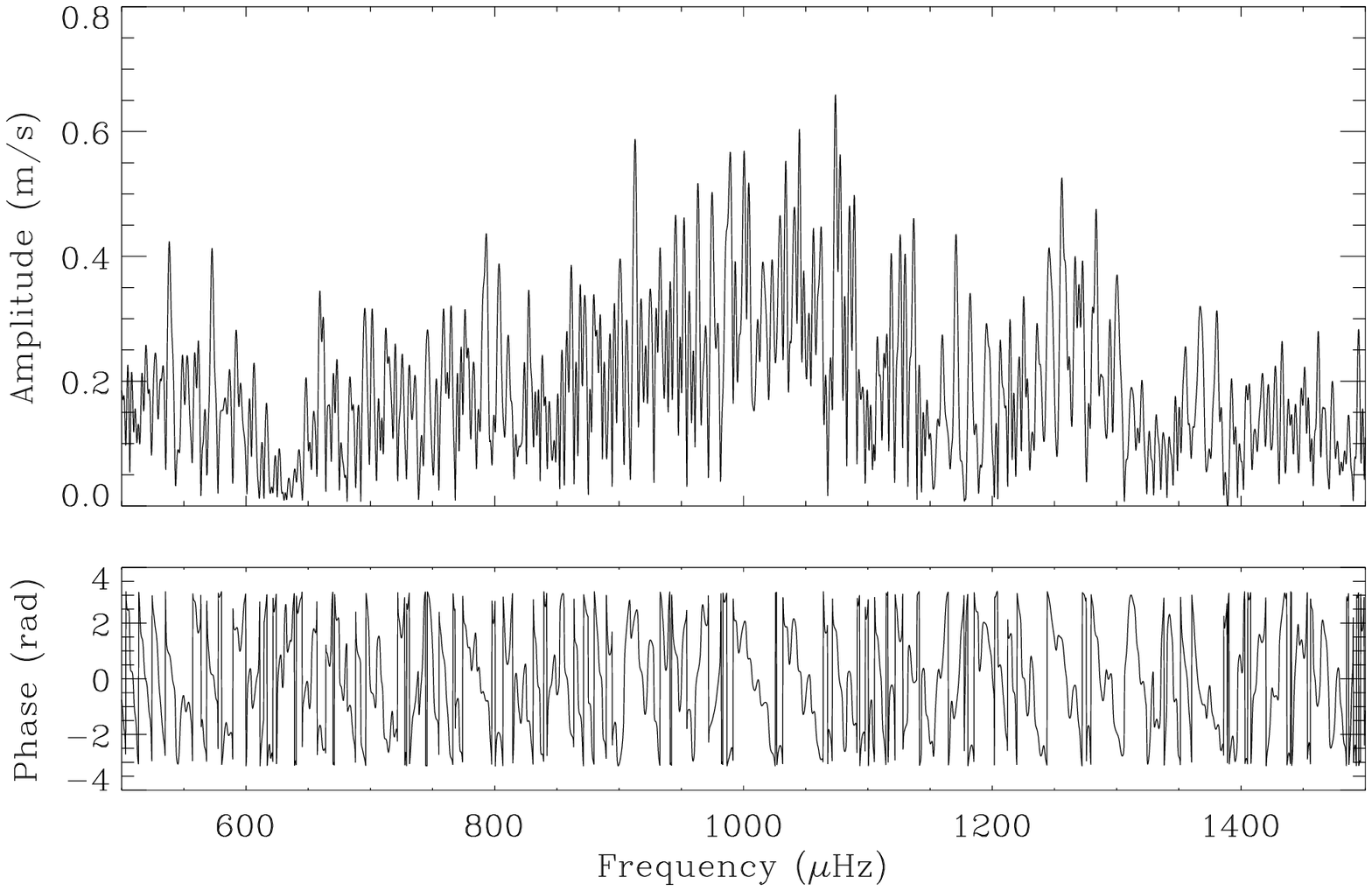}}
\caption{\label{fig.amp} Amplitude and phase of the UCLES velocity
measurements of \bhyi.}
\end{center}
\begin{center}
\mbox{\epsfxsize=0.95\textwidth\epsfbox{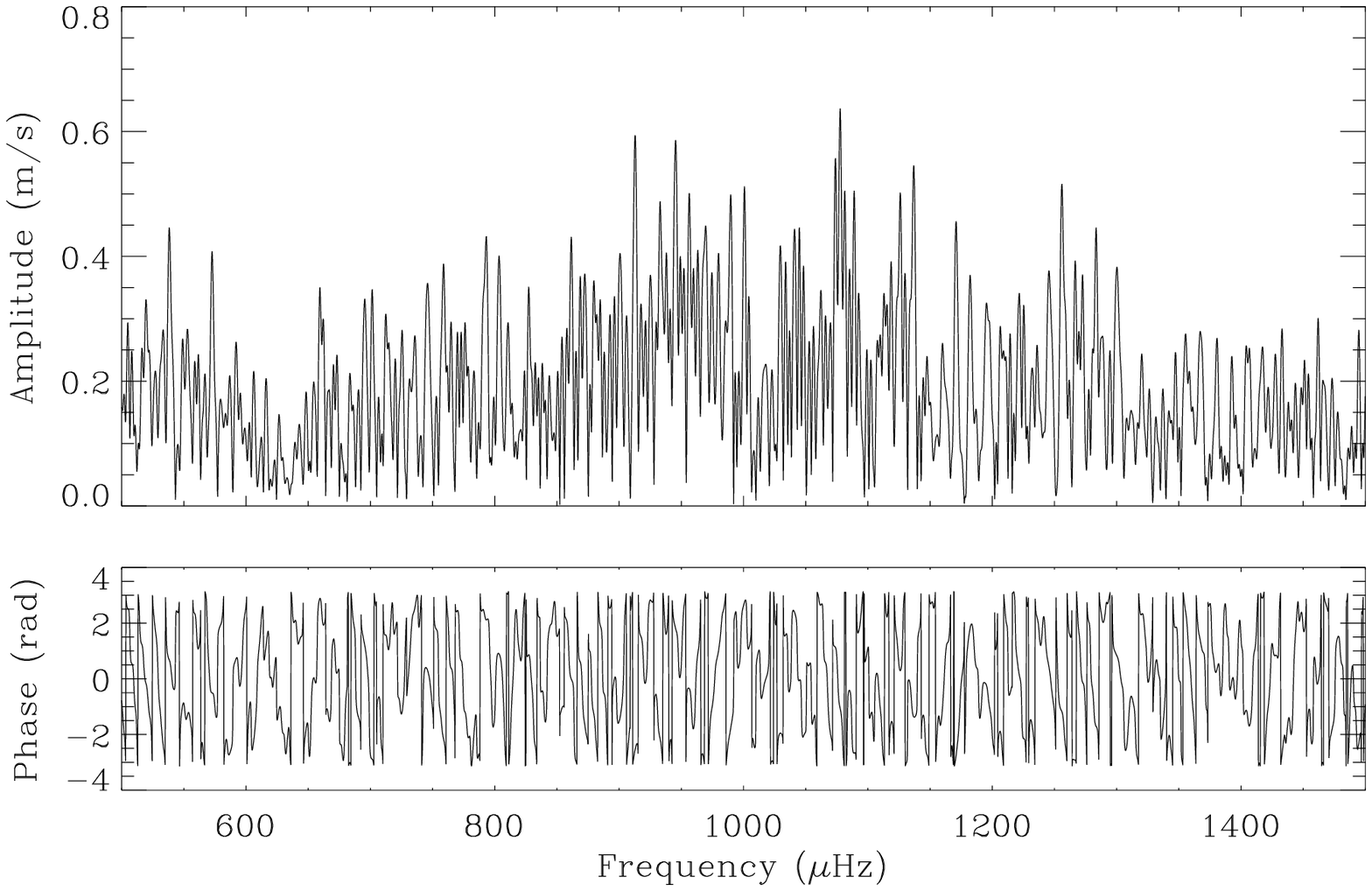}}
\caption{\label{fig.amp-quake} Same as Fig.\,2, but with part of
the velocity series removed (see text).}
\end{center}
\end{figure}

\section{Excess power}

We have compared the amount of excess power in the UCLES and CORALIE power
spectra, as follows.  We first converted both to power density by
multiplying the power (in (m\,s$^{-1})^2$) by the effective timespan of the
observations (see Appendix A.1 of Kjeldsen \& Bedding, 1995 and Section 5 of
Kjeldsen et al., 1999).  We calculated this effective timespan by
integrating under the (weighted) spectral window.  The timespans were
34.9\,hr for the UCLES data and 58.1\,hr for the CORALIE data.
We then calculated the mean power density in the range 700--1300\,$\mu$Hz
and subtracted the contribution from noise.  The latter was estimated from
the mean power density in the frequency range 1900--2200\,$\mu$Hz.  

The resulting values for the mean excess in power density are 0.0066 and
0.0054\,(m\,s$^{-1})^2/\mu$Hz, respectively for the UCLES and CORALIE
observations.  The agreement between these values (better than 20\%) is
good confirmation that both systems were measuring the same power excess.

We also checked the effect of deliberately shifting the relative time
stamps of the two series.  The amplitudes of the strongest ten or so peaks
in the combined power spectrum are maximized when this shift is zero and
become significantly reduced when the offset exceeds one minute.  This
gives strong evidence that highest peaks in the combined power spectrum
have a stellar origin.

\section{Starquakes~?} 

An unusual feature of the UCLES power spectrum of  (Fig.\,2 of Bedding et al.,
2001) is a region at the centre of the power excess (1.0\,mHz) where the
power is everywhere greater than zero.  One would expect the power spectrum
of a series of discrete long-lived sinusoids to fall to zero between the
peaks, not to produce a hump of entirely non-zero power.  This hump is more
obvious when seen in amplitude (square root of power), as shown here in the
upper panel of Fig.\,2.  There is a region $\sim$80\,\muHz\ wide
over which the amplitude stays well above zero.

Even more striking is the appearance of the phase (lower panel of Fig.\,2.)  The
phase is coherent over the region in question, which implies that this hump
arises from a specific location in the time domain.  That location can easily be
calculated from the phase, and turns out to correspond to $t_{\rm
event}=879,000$\,s (using the zero point of time as defined in the caption to
Fig.\,1).  We see in Fig.\,1 that this part of the velocity series shows clear
$\sim17$-minute variations, which must be the cause of the non-zero hump.  Could
this simply be the chance superposition of several modes that happened to be in
phase at that time~?  That would not produce the non-zero hump, as we have
verified with simulations.  As far as we can tell, such a hump can only be
produced by a short-lived variation that does not continue throughout the whole
time series.

We have re-calculated the Fourier spectrum after removing a 10000-s region
centred on $t_{\rm event}$ (73 points, indicated by the thick line in Fig.\,1).
As shown in Fig.\,3, this removes the non-zero hump and destroys the phase
coherence, confirming that we have identified the correct region of the time
series.

Without further evidence, we are left with a puzzle.  The UCLES velocities
appear to show a short-lived episode of large-amplitude oscillation that we are
tempted to describe as a starquake.  No similar events are seen in the CORALIE
data.  If confirmed as a feature of subgiants, this would be bad news for
asteroseismology, since they would make it harder to measure accurate mode
frequencies.

\end{document}